\documentclass{edp-conf}
\usepackage{graphicx}
\begin{document}

\TitreGlobal{SF2A 2002}

%%-----------------------------
%%      the top matter
%%-----------------------------
\title{Outflows in Galaxies and Damped Ly$\alpha$ Systems} 
\author{C\'eline P\'eroux}\address{Osservatorio Astronomico, Via
Tiepolo, Trieste, Italia}
%\author{Author2, B.}\address{...}
%\author{Author3, C.}\address{...}
%
\runningtitle{Outflows and DLAs}
\setcounter{page}{237}
% Keep this line, even if the page will be settled afterwards..
\index{C\'eline P\'eroux}
%\index{Author2, B.}
%\index{Author3, C.}
% Repeat the authors here, this will help to make the final index

\maketitle
\begin{abstract} 
Although quasar absorbers, and in particular Damped Lyman-alpha
systems (DLAs) have proven a valuable tool to study the early
Universe, their exact nature is so far poorly constrained. It has been
suggested that outflows in galaxies might account for at least part of
the DLA population. Observational evidences and models in support of
this hypothesis are reviewed, including recent observations of Lyman
Break Galaxies (LBGs). Observational counter-arguments and theoretical
limitations are also given. Finally, implications of such a model for
the environment of galaxies at high-redshifts are discussed.
\end{abstract}

%
%%-----------------------------
%%      your text
%%-----------------------------

\section{Quasar Absorbers}
Several observational techniques allow for the detection of
high-redshift galaxies, e.g. the Lyman-Break method, narrow band
Ly$\alpha$ imaging, and selection in other wavebands (radio, sub-mm,
etc). A complementary method is the detection of systems in absorption
in the spectra of background quasars. Since quasar are now detected to
$z\sim6$ (Fan et~al. 2001), quasar absorbers can be detected at very
early epochs. The quasar absorbers with the highest H~{\sc i} column
density are called Damped Lyman-$\alpha$ systems (hereafter DLAs)
since their line shapes show evidences for damping wings (Wolfe
et~al. 1986). These systems have proven powerful tools to study the
abundances of high-redshift objects (e.g. P\'eroux et~al. 2002a), in
terms of H~{\sc i} column density, ionic and molecular content
(e.g. P. Petitjean's contribution to these proceedings). The detection
of absorbers differs from more traditional techniques since it is
independent of the morphology and luminosity of these systems. DLAs in
particular can be used to trace the neutral gas (P\'eroux et~al. 2001
and reference herein) and metal abundances over cosmological scales
(e.g. P\'eroux et~al. 2002b). The limitations of studying galaxies
using absorbers is that only one-dimensional information is available
(with the exception of the growing body of transversal informations
now available from quasars groups and lensed system, e.g. B. Aracil's
contribution to these proceedings) and the fact that their exact
nature is unknown.

Several hypotheses have been proposed to explain DLAs, but imaging of
low redshift systems has shown that a variety of objects are at the
origin of the absorbers. Therefore the main questions are: {\it which
population of galaxies dominates DLAs at high redshift?  How does this
evolve with cosmological time?} Several approaches have been suggested
to answer these questions: i) comparing observations with models of
galaxy evolution (e.g. Boissier, P\'eroux \& Pettini, 2002); ii)
direct imaging of the systems giving rise to DLAs (e.g. Le Brun
et~al. 1997); iii) comparison with other types of high-redshift
galaxies (e.g. Bouch\'e \& Lowenthal 2000). It is the later which is
described here.

\section{Lyman-Break Galaxies}

Lyman-Break Galaxies (hereafter LBG) are selected using deep imaging
in 3 bands. This technique is particularly efficient to select star
forming galaxies at $z>2.5$. Several hundreds of these galaxies have
been found at $z\sim3$ (Steidel et~al. 1999). Their star formation
rate is typically between 1$-$100 M$_{\odot}$ yr$^{-1}$ and their
metallicities range from 1/10 to 1/2 solar. In addition, the velocity
offsets of the interstellar absorption lines and of the Ly$\alpha$
emission line relative to [O~{\sc iii}] and H$\beta$ indicate that
many of these objects possess high velocity outflows (100$-$1000 km
s$^{-1}$) (Pettini et~al. 2001). Such winds are important: they might
play a role in the feedback process required to regulate star
formation, they might be at the origin of metal pollution of the
interstellar medium (Ferrara et~al. 2000), and could create cavities
through which photons would escape and reionise the Universe at
high-redshift (Pettini et~al. 2001). Indeed, recent simulations shown
that outflows might be responsible for metal pollution without
destructing the filaments producing the Lyman-$\alpha$ forest (Theuns
et~al. 2002).

One particular LBG, MS 1512$-$cB58 ($z=2.72$) which has its flux
amplified by gravitational lensing due to a foreground cluster, has
been studied in details (Pettini et~al. 2000 \& 2002). Pettini
et~al. (2002) derived the chemical content of MS 1512$-$cB58 using 48
UV lines. They deduce that the metal enrichment took place some 300
Myr ago, indicating that the galaxy is rapidly transforming its
baryonic mass in stars (probably leading to the formation of a
galactic bulge or an elliptical). The Ly absorption of that galaxy can
be fitted with a damping profile with a hydrogen column density $N(HI)
= 7.5 \times 10^{20}$ cm$^{-2}$ (Pettini et~al. 2002), revealing an
asymmetric emission line. The complex structure of the Ly$\alpha$ line
is interpreted as the signature of photons scattered by outflowing
material.

\section{Are Outflows Responsible for DLAs?}

\subsection{Evidence in Favour}

Using geometrical arguments, Schaye (2001) has shown that at $z \sim
3$ the observed number density of DLAs ($dN/dz=0.20$) can be explained
by the comoving observed number density of LBGs ($0.016 \rm h^3 \rm
Mpc^{-3}$) assuming plausible shell's radius ($19 \rm h^{-1} \rm kpc$)
and luminosity ($<<L_*$). Other observational evidences have been
provided by Rauch et~al. (2002) who have studied Mg~{\sc ii} absorbers
along three adjacent lines of sight and find that the signature of
these systems is consistent with such an expanding shell.

Other authors have directly compared the properties of galaxies with
those of DLAs. Heckman et~al. (2001) used Na~{\sc i} lines to study
local near-IR bright galaxies. From the Na~{\sc i} equivalent width,
they deduce a column density of about $\sim 10^{21}$ cm$^{-2}$ in the
outflows. Therefore, if a luminous quasar were located behind such
line of sight, its spectrum would contain a damping profile as well as
low-ionisation metals spanning velocity range up to few km s$^{-1}$,
similar to what is observed in high redshift DLAs. Moller
et~al. (2002) compared the properties of 3 spectroscopically confirmed
HST-imaged DLAs ($z_{\rm abs} =$ 1.92, 2.81 \& 3.15) with those of
LBGs. They find that the half-light radius, the radial profile, the
optical-to-near-IR colour, the morphology, and the Ly$\alpha$ emission
equivalent width and velocity structure match well those of LBGs. They
thus conclude that a least part of the DLAs could be linked to LBGs.

\subsection{Evidence Against}

Nevertheless, three important observational characteristics appear
extremely different in DLAs and in LBGs, as pointed by Pettini
(2001). These are:

\begin{enumerate}
\item{an important difference in their inferred star formation rates
(around 50 M$_{\odot}$ yr$^{-1}$ for LBGs; less than 10 M$_{\odot}$
yr$^{-1}$ for DLAs)}
\item{a difference in terms of metallicities (1/3 solar in LBGs; 1/20 solar in DLAs)}
\item{a difference in the velocity of the outflows (around 500 km
s$^{-1}$ for LBGs; while less than 200 km s$^{-1}$ for DLAs)}
\end{enumerate}

It has been suggested (Theuns et~al. 2001) that LBGs are more metal
rich and thus more dusty, occulting their presence along the line of
sight to a background quasar. Schaye (2001) also proposed that, for
geometrical reasons, DLAs preferentially occur in the external part of
the outflows with higher cross-section and where the star formation
rate is lower, thus explaining the observed discrepancies. Finally,
Pettini (2001) argue that LBGs are quickly transforming their baryonic
mass in stars, while DLAs have a slower star formation rate,
suggesting that the two populations probe different types of galaxies.

\section{Summary}

Recent observations have shown that outflows might be common amongst
high-redshift galaxies. The nature of quasar absorbers, extensively
studied in the past years, is not as yet strongly constrained. It is
suggested that at least part of the DLA population is due to the
outflows in galaxies, although some evidence seems to disfavour such
an hypothesis. Therefore these may have important consequences for
regulating star formation, polluting the intergalactic environment
with metals and indirectly to the reionisation of the Universe.

%\vspace{.5cm}

{\bf Acknowledgments:} I am grateful to Tom Theuns for valuable
discussions on this topic.

%\begin{figure}[h]
%   \centering
%  If you have a figure, remove the comments % before...
%   \includegraphics[width=9cm]{author1_fig1.ps}
%      \caption{ My caption here..}
%       \label{figure_mafig}
%   \end{figure}

%%-----------------------------
%%      your bibliography
%%-----------------------------

\end{document}